\documentstyle[prl,aps,epsfig]{revtex}

\tightenlines

\begin{document}
\title{Phase Coherence and Control of Stored Photonic Information}
\author{A.~Mair$^1$, J.~Hager$^1$,
D.~F.~Phillips$^1$,   R.~L.~Walsworth$^1$, and
M.~D.~Lukin$^2$}
\address{$^1$ Harvard-Smithsonian Center for Astrophysics,
                Cambridge, MA~~02138, \\ $^2$
ITAMP, Harvard-Smithsonian Center for Astrophysics,
                Cambridge, MA~~02138 }

\date{\today}
\maketitle

\begin{abstract}
  We report the demonstration of phase coherence and control for the
  recently developed ``light storage'' technique. Specifically, we use
  a pulsed magnetic field to vary the phase of atomic spin excitations
  which result from the deceleration and storing of a light pulse in
  warm Rb vapor.  We then convert the spin excitations back into light
  and detect the resultant phase shift in an optical interferometric
  measurement.  The coherent storage of photon states in matter is
  essential for the practical realization of many basic concepts in
  quantum information processing.
\end{abstract}


The realization of scalable quantum networks for long-distance quantum
communication and quantum computation requires the use of photons as
quantum information carriers and matter (e.g., spins) as quantum
memory elements \cite{qi}.  For example, intermediate memory nodes
will be essential for quantum communication over lossy photonic
channels \cite{enk,briegel}. When successfully implemented, such a
technique will facilitate secure transmission of secret messages over
long distances \cite{qcrypt1,qcrypt2}. Likewise, quantum memory
elements linked by light are desirable for scalable quantum
computation \cite{divin}.

The quantum carrier/memory interface is a key component that should be
capable of reversibly transferring quantum states between light pulses
and long-lived matter states. In contrast to ordinary ``destructive''
techniques which convert light into, e.g., electrical signals by
photoabsorption, a quantum memory should be free from dissipation and,
most importantly, should preserve phase coherence in the process of
information transfer to and from the carrier.

Recently, proposals have been made to accomplish the quantum transfer
of photon states to individual atoms \cite{QED} and atomic ensembles
\cite{storageTheory,polaritons}.  Active experimental efforts toward
the realization of these ideas are currently under way
\cite{kimble2,rempe2}.  In particular, recent experiments employing
cold Na atoms \cite{haustorage} and warm Rb vapor \cite{lightstorage}
demonstrated the basic principle of the ``storage of light'' in atomic
ensembles by the dynamic and reversible reduction of the light pulse
group velocity to zero.

In this Letter we present the first experimental evidence that this
light storage technique is phase coherent.  Although anticipated from
theoretical predictions \cite{storageTheory,polaritons}, this
essential feature of a quantum memory for light has not been verified
experimentally up to now.  In addition, we demonstrate that the phase
of the stored coherence can be accurately manipulated during the
storage interval and then mapped coherently onto the released light
pulse. These results show that the present technique should be
suitable for applications in quantum information processing. For
example, a specific ``quantum repeater'' protocol, which allows for
scalable quantum communication over very long distances using this
technique, has already been proposed \cite{duan}.  Scalable quantum
computation using atomic ensembles coupled by light has also been
suggested \cite{lukin01}.

\begin{figure}[ht]
\centerline{\epsfig{file=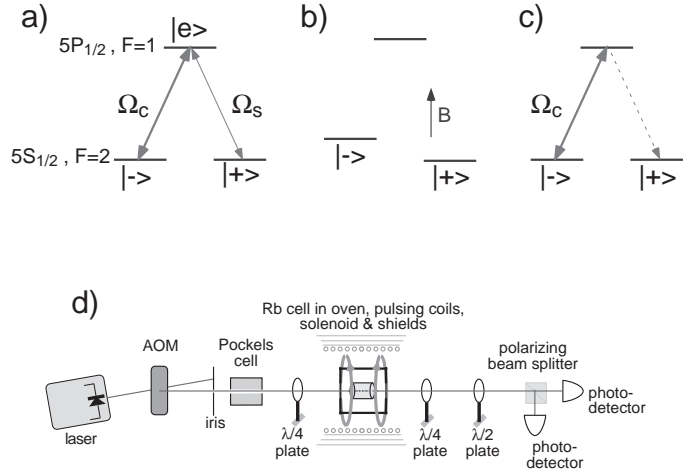,width=9.0cm}}
\vspace*{2ex}
\caption{Schematic of system for ``light storage'' and phase
manipulation.  (a) $\Lambda$-type configuration of $^{87}$Rb atomic
states resonantly coupled to a control field ($\Omega_c$) and a signal
field ($\Omega_s$).  (b) Phase manipulation of Zeeman coherence using
magnetic field.  (c) Retrieval of stored photonic excitation using
control field.  (d) Schematic of the experimental setup.
Approximately 5 torr of helium buffer gas was used to keep Rb atoms in
the laser beam for approximately 200 $\mu$s and thus to maintain long
storage times.  Note the addition since our previous experiment
\protect\cite{lightstorage} of Helmholtz coils to pulse the magnetic
field at the Rb cell.
}
\label{f.setup}
\end{figure}

The light storage technique is based on the phenomenon of
Electromagnetically Induced Transparency (EIT) \cite{rev}, in which an
external optical field (the ``control field'') is used to make an
otherwise opaque medium transparent near an atomic resonance. A
second, weak optical field (the ``signal field'') at an appropriate
frequency and polarization can then propagate without dissipation and
loss but with a substantially reduced group velocity
\cite{hau99,kash99,budker99}.  The present experiment can be
understood by considering a $\Lambda$ configuration of atomic states
coupled by two optical fields (see Fig.~\ref{f.setup}a).  Here, the
control field (Rabi-frequency $\Omega_c$) and signal field
($\Omega_s$) are left and right circularly polarized light ($\sigma^+$
and $\sigma^-$).  Via Raman transitions, these light fields create in
the atomic ensemble a coherent antisymmetric superposition of a pair
of ground-state Zeeman sublevels ($|-\rangle, |+\rangle$) which have
magnetic quantum numbers differing by two.  Associated with the
reduced group velocity of the signal field is a considerable spatial
compression which allows a signal pulse to be almost completely
localized in the atomic medium.

Inside the medium the light pulse propagates together with the
ensemble Zeeman coherence, like a wave of flipped spins.  The
propagation of the coupled light and spin excitations can be
efficiently described in terms of a quasi-particle, the ``dark-state
polariton'' \cite{polaritons}, which is a coherent superposition of
photonic and spin-wave contributions. In order to store the state of a
pulse of signal light, one smoothly turns off the control field,
causing the dark-state polariton to be adiabatically converted into a
purely atomic spin excitation which is confined in the vapor cell.

The phase of an atomic Zeeman coherence can be easily manipulated using an
external magnetic field. If a pulsed magnetic field, $B_{z}(t)$, is applied
in a direction parallel to that of the light propagation (the quantization
or $\hat{z}$ axis), then the Zeeman
sublevels are differentially shifted in energy (see
Fig.~\ref{f.setup}b), producing a phase shift in
the Zeeman coherence given by:
\begin{equation}
\Phi = (g_+- g_-) {\mu_B \over \hbar} \int_0^T B(t') dt'.
\label{e.phase}
\end{equation}
Here $g_\pm$ are Lande factors corresponding to the Zeeman states
$|\pm\rangle$ and $T$ is the time during which the magnetic field is
applied. Once the magnetic field is removed, the dark-state polariton
can be adiabatically restored to a photonic excitation by turning the
control field ($\Omega_c$) back on (see Fig.~\ref{f.setup}c).  The
phase-shift $\Phi$ of the atomic coherence is thus transfered to a
phase difference between the control field and the reconstituted
signal pulse emitted from the sample.

Similar to our previous work \cite{lightstorage} we performed
light-storage experiments in atomic Rb vapor at temperatures of
$\sim70-90 \ {}^o$C, corresponding to atomic densities of $\sim
10^{11}-10^{12}$ cm$^{-3}$.  Under these conditions the 4 cm-long
sample cell was normally opaque for a single, weak optical field near
the Rb $D_1$ resonance ($\simeq$ 795 nm).  To create the conditions
for EIT, we derived control and signal beams from the output of the
same extended cavity diode laser by carefully controlling the light
polarization as illustrated in the experimental schematic shown in
Fig.~\ref{f.setup}d.  For the data presented here we employed the
$5^{2}S_{1/2}, F=2 \rightarrow 5^{2}P_{1/2}, F=1$ (i.e., $D_1$)
transition in $^{87}$Rb.  The control field ($\sigma^+$ polarization
light) was always much stronger than the $\sigma^-$ signal field
($\Omega_c \gg \Omega_s$); hence most of the relevant atoms were in
the highest angular momentum ground-state sublevel.  In this case the
states $|-\rangle, |+\rangle$ of the simplified 3-level model
correspond, respectively, to $|F=2,M_F = 0\rangle$ and $|F=2,M_F =
+2\rangle$.  We circularly-polarized the input laser beam to create
the $\sigma^+$ control field, and then collimated and focused the beam
to a diameter of about 1 mm as it passed through the Rb vapor cell. By
using a Pockels cell we slightly rotated the polarization of the input
light to create a weak pulse of $\sigma^-$ light, which served as the
signal field. Input peak powers were about 1 mW and 100 $\mu$W for the
$\sigma^+$ and $\sigma^-$ components. Peak power in the control field
provided a transparency bandwidth for the signal pulse \cite{sau} of
approximately 40 kHz.

Additional polarizing optics after the atomic vapor cell allowed us to
monitor the output signal light pulse (see Fig.~\ref{f.setup}d).  A
$\lambda/4$ waveplate converted the two circular polarizations into
linear polarizations, followed by a $\lambda/2$ waveplate, a
polarizing beam splitter, and two photodetectors to provide two
detection channels -- nominally for the control and signal fields.  To
form an interferometer for the two fields, we adjusted the $\lambda/2$
plate such that a small fraction ($<$ 10$\%$) of the control field was
mixed into the signal detection channel with approximately the same
steady-state magnitude as the peak of the released signal pulse.

To manipulate the phase of the atomic Zeeman coherence, we carefully
controlled the magnetic field at the Rb cell.  To this end, we
employed three layers of high permeability magnetic shields to screen
out external magnetic fields.  Additionally, we used two separate
coils to generate homogeneous magnetic fields along the propagation
direction of the optical beam.  First, a precision solenoid produced a
static magnetic field of less than 1 mG to offset the remnant field
unscreened by the shields.  Second, a Helmholtz coil pair generated
the pulsed field to manipulate the Rb Zeeman coherence, thereby
inducing phase shifts in the stored light.  We typically applied
current pulses to the Helmholtz pair of 10 to 30 $\mu$s duration, with
peak magnetic fields as large as 100 mG \cite{nonlinNote}.

\begin{figure}[ht]
\centerline{\epsfig{file=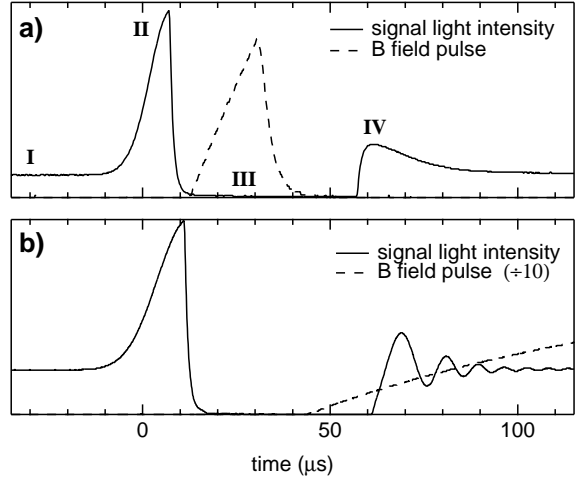,width=9.0cm}}
\vspace*{2ex}
\caption{Examples of ``stored light'' with interferometric
detection and a pulsed magnetic field. The dashed line indicates the
pulsed magnetic field; the solid line is the measured signal field. A
small portion of the control field ($\sigma^+$) was mixed into the
signal field ($\sigma^-$) detection channel by a slight rotation of
the $\lambda/2$ waveplate after the Rb cell. Thus the relative phase
$\Phi$ between the outgoing control field and the released signal
pulse could be determined from the degree of constructive and/or
destructive interference between these two fields.  (a) Pulse is
applied during the storage interval.  In the example shown here, $\Phi
\simeq 4 \, \pi$, leading to constructive interference at the region
marked IV.  (The significance of the other three regions is explained
in the main text.)  (b) Pulse is on during the retrieval procedure.
The frequency of the output signal pulse becomes shifted, resulting in
interferometric beating in the measurement.
}
\label{f.sample}
\end{figure}

Fig.~\ref{f.sample}a shows an example of stored light observed with an
applied magnetic field pulse which created an approximately $\Phi =
4\pi$ phase shift between the signal pulse and the control field. With
the $\sigma^+$ control field dressing the atoms, we measured a small
background signal due to the control field being slightly mixed into
the signal detection channel via the slightly-rotated $\lambda/2$
plate (region I of Fig.~\ref{f.sample}a). We then applied a $\sigma^-$
signal pulse of duration $\sim$ 20 $\mu$s.  Upon entrance into the
sample cell the signal pulse was spatially compressed by more than
five orders of magnitude in the Rb vapor due to the reduction in group
velocity, as estimated from the observed delay in pulse propagation.
With the peak amplitude of the signal pulse inside the cell, we ramped
down the control field, effectively storing the pulse in an ensemble
Zeeman coherence in the Rb vapor.  Note that approximately half the
signal pulse had already left the cell before the beginning of the
storage procedure (region II) \cite{optimized}. This front part of the
signal pulse was not affected by the phase manipulations and was used
as a reference in the experiment.  With the remainder of the signal
pulse stored in the vapor cell, we applied a pulsed current to the
Helmholtz coils to induce a controllable phase shift in the $^{87}$Rb
ensemble Zeeman coherence. After the pulsed magnetic field, we turned
the control field back on and the signal light pulse was
reconstituted, emitted from the sample, and observed
interferometrically (region IV). The integrated magnetic field pulse
in the example shown in Fig.~\ref{f.sample}a was 1.5 gauss $\cdot
\mu$s, producing a signal/control field phase shift of about $4 \,
\pi$: i.e., constructive interference between the two optical fields.

Note that for the accurate, coherent control of the phase it is
essential to apply the magnetic field only during the storage
interval.  For example, the peak magnetic field in
Fig.~\ref{f.sample}a corresponds to a Zeeman frequency shift which is
at least four times larger than the EIT transparency window.  If such
a field is applied before the storage is complete, the medium becomes
completely opaque and no signal pulse is recovered. Alternatively, if
the magnetic field is left on during the retrieval procedure, the
light is retrieved at a different frequency resulting in a modulated
interferometric signal (Fig.~\ref{f.sample}b). (Only the control field
is applied to the atoms during release of the stored signal pulse,
which obviates the need for degenerate Zeeman levels to maintain EIT.)

\begin{figure}[ht]
\centerline{\epsfig{file=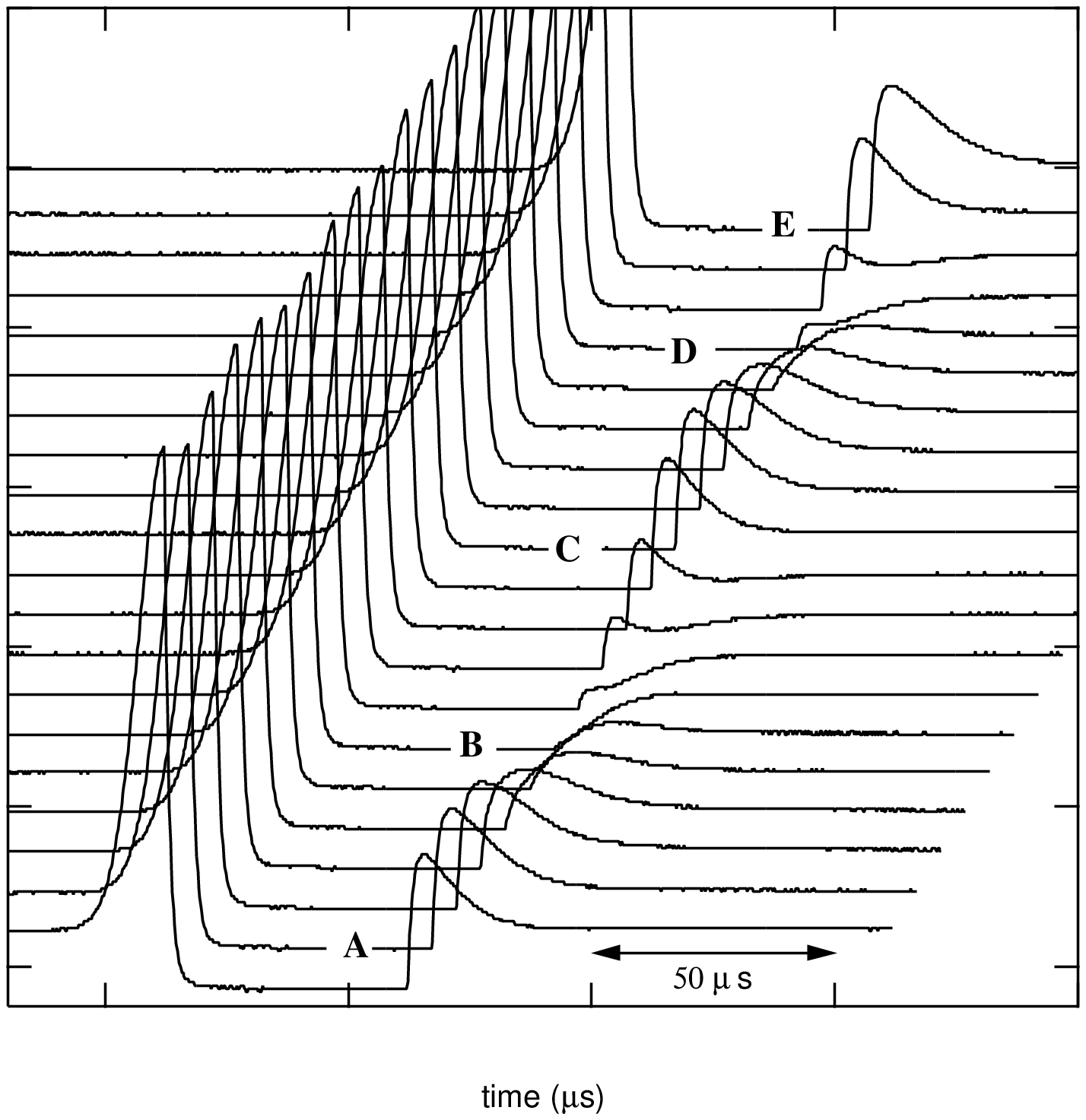,width=9.0cm}}
\vspace*{2ex}
\caption{Results of interferometric measurements of released
photonic excitations: twenty light storage experiments similar to that
shown in Figure~\protect\ref{f.sample}. The magnetic field was pulsed
during the storage interval with increasing strength from trace A to E
such that the accumulated phase difference between the output signal
pulse and the control field varied from approximately 0 to $4\pi$.
Note: there is a small phase offset at zero pulsed magnetic field,
caused by the Pockels cell.
}
\label{f.waterfall}
\end{figure}

By adjusting the applied magnetic field pulse during the storage, we
could easily modify the phase $\Phi$ of the coherent excitation.
Figure~\ref{f.waterfall} shows twenty stored light experiments for
which we increased the Zeeman phase shift by approximately $0.2 \,
\pi$ for each successive run.  Trace A in Fig.~\ref{f.waterfall} shows
the result for $\Phi \simeq$ 0 and hence maximum constructive
interference between the output signal light and the control field.
As we increased the pulsed magnetic field to change the phase by
$\pi$, we observed destructive interference (e.g., trace B).  As we
increased the pulsed magnetic field still further, we alternatively
observed constructive and destructive interference as expected at
$\Phi \simeq 2 \, \pi, 3 \, \pi, 4 \, \pi$, etc.\ (traces C-E in
Fig.~\ref{f.waterfall}).  (Note: Fig.~\ref{f.sample}a and trace E in
Fig.~\ref{f.waterfall} show the same data.)  We observed up to 10
periods of phase accumulation (i.e., $\Phi \simeq 20 \, \pi$) without
loss of coherence, limited by the strength of our pulsed magnetic
field power supply rather than non-linear shifts in the Rb Zeeman
levels at higher fields or magnetic field inhomogeneity.

In conclusion, we have demonstrated that the recently developed
``light storage'' technique \cite{lightstorage} is phase coherent. In
so doing we have performed accurate, coherent manipulation of
information that is stored in an atomic spin coherence and then
transfered back into light and released. The present work may
facilitate the practical implementation of scalable quantum repeater
protocols \cite{duan} and new techniques for quantum computation
\cite{lukin01}.  It is a pleasure to thank M.~Fleischhauer and
S.~Yelin for many fruitful ideas and collaboration on theoretical
aspects of this work.  We also thank M.~O.~Scully for many stimulating
discussions. This work was partially supported by the National Science
Foundation through the grant to ITAMP, the Office of Naval Research,
and NASA.

\newpage

\begin{thebibliography}{99}
\bibitem{nonlinNote} At fields of 100 mG, differential shifts in the Zeeman
frequencies between different levels and
deviations from the linear approximation are $< 2$ Hz and $3 \times
10^{-5}$, respectively, which is negligible in
the present experiment.
\bibitem{optimized} We estimate that a signal pulse storage efficiency
of better than 90$\%$ should be possible in warm vapor with an
optimized system.
\end{thebibliography}
\end{document}